# Ultrafast graphene-based broadband THz detector

**Martin Mittendorff**[1,2], **Stephan Winnerl**[1], **Josef Kamann**[3], **Jonathan Eroms**[3], **Dieter Weiss**[3], **Harald Schneider**[1], **Manfred Helm**[1,2]

[1] Helmholtz-Zentrum Dresden-Rossendorf, P.O. Box 510119, 01314 Dresden, Germany

[2] Technische Universität Dresden, 01062 Dresden, Germany

[3] Universität Regensburg, 93040 Regensburg, Germany

Email: m.mittendorff@hzdr.de

Abstract:

We present an ultrafast graphene-based detector, working in the THz range at room temperature. A logarithmic-periodic antenna is coupled to a graphene flake that is produced by exfoliation on $SiO_2$. The detector was characterized with the free-electron laser FELBE for wavelengths from 8 µm to 220 µm. The detector rise time is 50 ps in the wavelength range from 30 µm to 220 µm. Autocorrelation measurements exploiting the nonlinear photocurrent response at high intensities reveal an intrinsic response time below 10 ps. This detector has a high potential for characterizing temporal overlaps, e. g. in two-color pump-probe experiments.

Since the first graphene flakes have been produced and investigated by Novoselov et al. [1] the development of optoelectronic graphene-based devices has been extremely fast. The special band structure of graphene [2], where conduction and valence band touch each other in the Dirac point, provides nearly constant photon absorption for the range from the visible light down to far infrared (FIR) radiation [3]. Combined with extremely high carrier mobility [4] and fast carrier relaxation [5] this enables one to realize ultrafast optoelectronic devices for an extremely wide spectral range. Detectors for near-infrared radiation (NIR), which can be used for telecommunication purposes with frequencies of up to 16 GHz, have been demonstrated [6]. In these devices a photocurrent is induced by an asymmetry in the detector, which is provided by two different metals serving as contacts to the flake. Recently detectors based on graphene field-effect transistors have been developed for THz frequencies. On the one hand a high-speed bolometer working at helium temperature has been demonstrated [7], on the other hand a rectifying transistor operating at room temperature has been reported [8]. For the latter only cw detection has been shown. The detector presented in this work combines room temperature operation with ultrafast response, including the potential to serve as an ultra-broadband detector up to the visible range.

The active material of our device is a graphene flake with a size of ~10 µm x 10 µm, produced by mechanical exfoliation of natural graphite onto a 300 nm thick layer of $SiO_2$ on silicon. The single-layer behavior of the flake is verified by Raman spectroscopy. Broadband THz response is achieved by processing a logarithmic-periodic antenna [9] onto the graphene flake by means of electron-beam lithography. The center of the antenna is structured into an interdigitated comb (see Fig. 1), which provides the contact to the graphene flake. The outer diameter of the antenna is 1 mm, which limits the maximum wavelength of the detected radiation. Wavelengths that are smaller than the shortest elements of the antenna (~10 µm) cannot efficiently couple to the antenna. Those wavelengths, however, can be focused to a spot of a size comparable to the graphene flake and thereby couple to the flake directly. For contacting the antenna to a coaxial cable (50 Ω with SMA connector) two contact pads with the size of 500 µm x 500 µm are connected via 20 µm thick and 500 µm long strip lines to the outermost part of the antenna.

Two different types of detectors were tested. A first set of detectors was produced on substrates of undoped Si with a resistivity of 10 kΩ cm underneath the $SiO_2$. For this set the antenna metallization consists of 60 nm Au on top of 5 nm Ti. The Si for the second set is heavily p-doped with a resistivity of about 5 mΩ cm. In this case two different metals were used for the two arms of the antenna. One arm consists of a 60 nm thick layer of Pd, the second one of a 20 nm thick layer of Ti with 40 nm Au on top. The intention to use two different metals is to break the symmetry of the device and thereby increase the photocurrent [6]. The combination of graphene flakes with a logarithmic-periodic antenna results in a device resistance in the range of several hundred Ω, the device on high-resistive substrate used for the measurements had a resistance of 280 Ω.

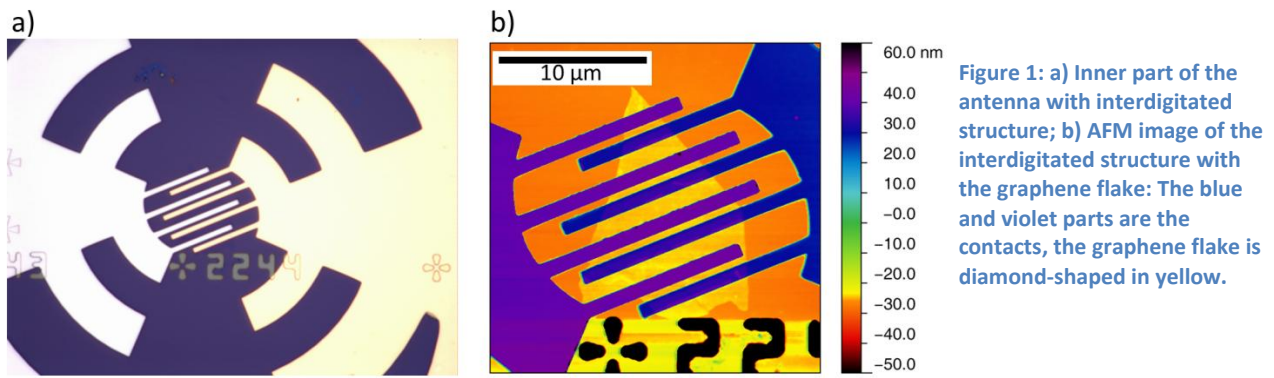

Figure 1: a) Inner part of the antenna with interdigitated structure; b) AFM image of the interdigitated structure with the graphene flake: The blue and violet parts are the contacts, the graphene flake is diamond-shaped in yellow.

The devices were characterized using the free-electron laser (FEL) at Dresden-Rossendorf (FELBE). It provides infrared and THz pulses in the wavelength range from 5 µm to 250 µm with a repetition rate of 13 MHz. The average power for our experiments was between 1 mW and 1 W which corresponds to pulse energies between 80 pJ and 80 nJ. The FEL beam was focused on the devices by off-axis parabolic mirrors with focal lengths of 50 mm to 100 mm. The rise time of the devices was measured with a sampling oscilloscope with a bandwidth of 30 GHz. A high frequency amplifier was used to increase the amplitude of the signals and the signal-to-noise ratio. Additionally we performed measurements of the time-integrated photocurrent with a lock-in amplifier. In this case the FEL beam was modulated by a mechanical chopper. This technique allowed us to measure very small photocurrents to investigate the linear regime of the detector. A bias voltage could be applied to the detectors for all measurements using a bias tee. The lock-in technique was furthermore used in autocorrelation measurements. Each FEL pulse was split up into two pulses and subsequently recombined collinearly by a pair of Si wafers. The time delay between the pulses was varied by a mechanical delay stage.

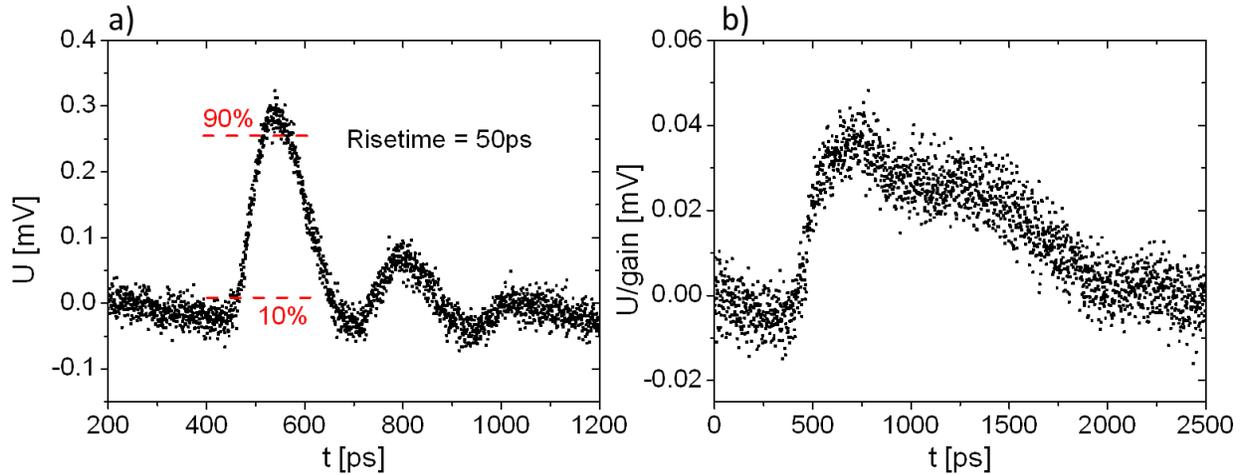

Figure 2: a) Signal shapes measured with a device on high-resistive substrate at a wavelength of 68 µm and an average FEL power of 72mW. For this measurement neither a bias voltage was applied, nor was the high speed amplifier used. b) Signal shape measured with a device on low-resistive substrate. A bias voltage of 0.1 V was applied, additionally the high frequency amplifier was employed, for a better comparison the voltage was divided by the gain (50) of the amplifier.

In Fig. 2a) the fast electric response to an FEL pulse of a detector fabricated on the high-resistive substrate is shown. The rise-time of the signal is (50 ± 10) ps. We assume the second pulse to be caused by signal reflections at the contact to the coaxial signal cable. By applying a bias voltage of 100 mV the amplitude of the measured signal could be increased above 1mV. In combination with a high frequency amplifier the signal amplitude reached 50 mV. Changing the substrate material from high-resistive Si to low-resistive Si strongly decreases the high speed performance of the device. For the low-resistive Si the rise time is in the range of 100 ps while the signal pulse length is increased to above 1 ns (see Fig. 2 b)). The amplitude of the signal is strongly decreased compared to the signal of the device on high-resistive substrate, while the area under the pulse stays roughly constant. This is consistent with the observed time-integrated photocurrent, which is similar in both cases. For both types of detectors we could not observe any fast signals at wavelengths below 20 µm while the time-integrated photocurrent remained constant (see Fig. 3 b)).

The saturation behavior of all devices (with different substrates and metallizations) was determined by measuring the photocurrent in dependence of the FEL pulse energy. The measured data were fitted by the expression

$$I_{phot} \propto \frac{E/E_{sat}}{1+E/E_{sat}}. \tag{1}$$

From the fit the saturation pulse energy $E_{sat}$ was extracted [10] (see Fig. 3 a) and 3c)). At this pulse energy the photocurrent $I_{phot}$ is suppressed by 50% compared to the fictitious photocurrent without saturation.

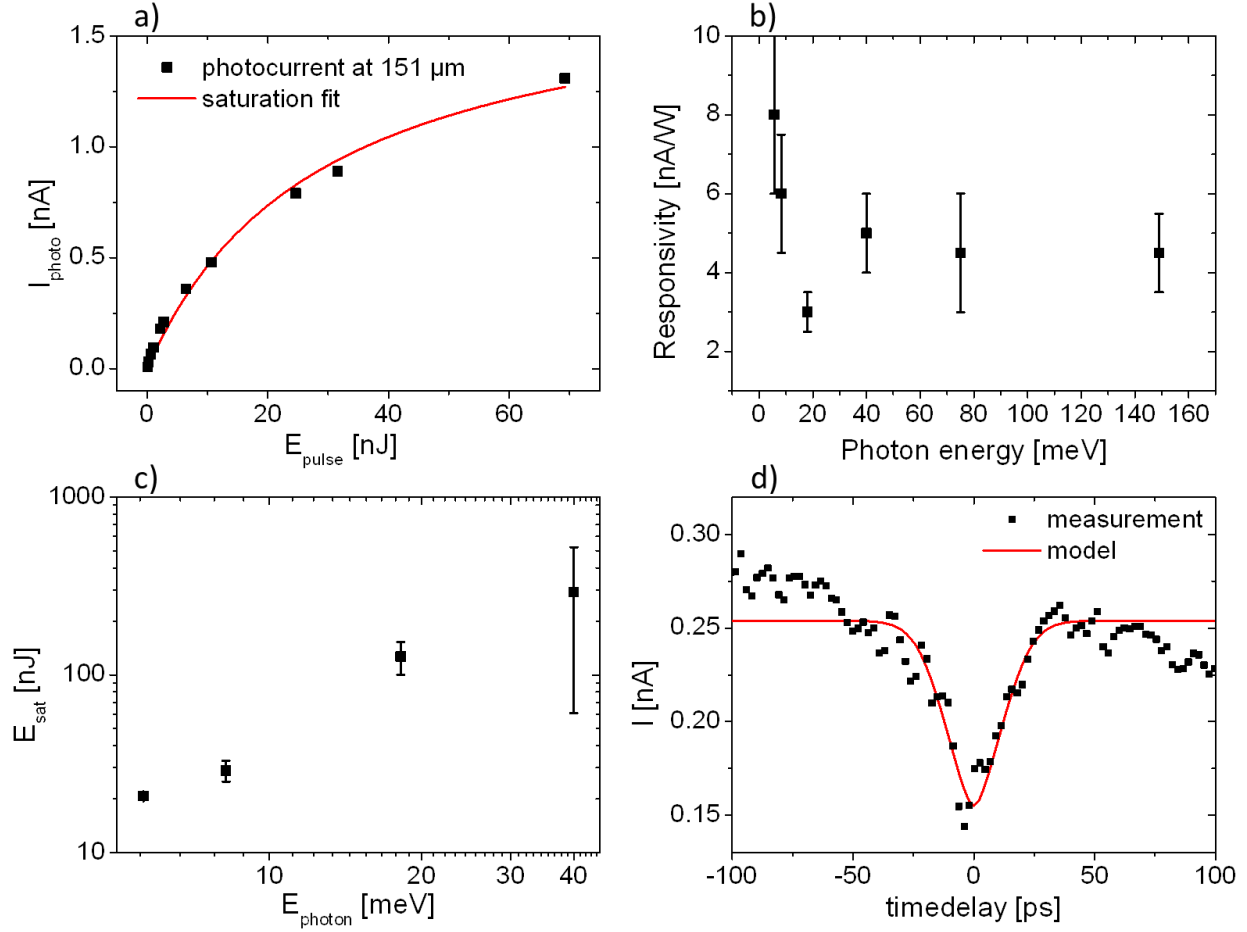

Figure 3: a) Time-integrated photocurrent as a function of the FEL pulse energy at 151 µm. b) Responsivity measured with the lock-in amplifier at different photon energies. c) Saturation pulse energy $E_{sat}$ as a function of the photon energy. d) Autocorrelation signal at a wavelength of 42 µm and a FEL-pulse duration of 7 ps.

The saturation energy increases with decreasing wavelength (cf. Fig. 3 c)), which is consistent with the increasing density of states for higher energies. For the measurement at a wavelength of 31 µm ($E_{photon}$ = 40 meV) the error bar is very large due to the nearly linear power dependence of the photocurrent and the associated high tolerance of the fit parameters. For even higher photon energies (75 meV & 149 meV) we could not observe any saturation and the photocurrent was increasing linearly with the FEL pulse energy. The responsivity was in the range of 5 nA/W for all wavelengths (cf. Fig. 3 b)). With the strong saturation at long wavelengths the devices can serve as nonlinear detectors in autocorrelation setups [11]. In this configuration the temporal resolution is not limited by the electronic circuit, but by the intrinsic mechanism resulting in the saturated photocurrent response. In Fig. 3 d) the measured autocorrelation signal obtained at a wavelength of 42 µm ($E_{photon}$ = 30 meV) as well as calculated data are displayed. In the experiment the time delay was varied continuously resulting in an intensity-autocorrelation trace without interference fringes. The autocorrelation trace exhibits a dip at zero time delay due to the saturated photocurrent response of the detector. The calculation was performed with a Gaussian FEL pulse (full width at half maximum: 7 ps [12]) with replicas, caused by the Si beam splitter and combiner. The replicas were delayed by multiples of 7 ps with respect to the initial pulse, i.e. the pulses merge to one prolonged pulse. The autocorrelation signal was calculated based on the nonlinear current described by Eq. 1. The fit to the experimental data was performed with only one adjustable parameter for scaling of the absolute pulse energy. The shapes of the dips of the experimental and calculated curve agree well (cf. Fig. 3d)), indicating that the intrinsic response time of the detector is similar or shorter than the duration of the FEL pulse. The intrinsic response time of a few ps is consistent with results of pump-probe experiments at room temperature [5].

The coupling of the antenna to the graphene flake was verified in measurements with different rotation angles between the orientation of the antenna and the polarization of the radiation. In Fig. 4 the results of these measurements are plotted for two different wavelengths. At 151 µm the angle of the maximum photocurrent is rotated by ~70 ° compared to the maximum for 68 µm. As radiation of different wavelength couples to different parts of the antenna exhibiting resonant dimensions, this proves the operation of the antenna.

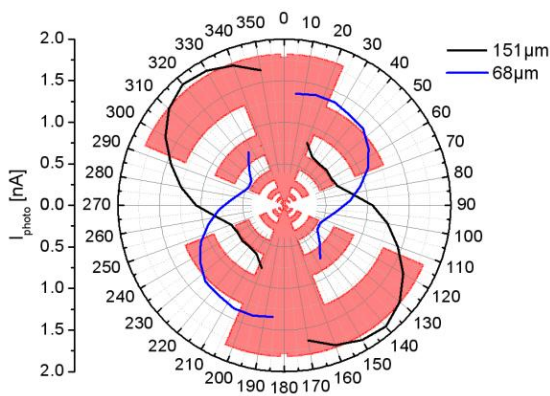

**Figure 4: Polarization dependent photocurrent for two different wavelengths, in the background the orientation of the antenna is depicted.**

Finally it was tested whether a pure antenna on the Si/SiO$_2$ substrate without a graphene flake also produces current signals under THz irradiation. No signals were found with the sampling oscilloscope, hence all observed fast signals of the previous measurements can clearly be attributed to the graphene flake. In contrast, small photocurrents were observed with a lock-in amplifier for the sample without graphene flake, which we assume to be caused by parasitic effects. These signals were about 10 times smaller than the signals of the devices with graphene flake.

The detection mechanisms leading to a photocurrent in graphene induced by a NIR laser have been discussed in detail by Freitag et al. [13]. They distinguish three main effects contributing to the current, namely a photovoltaic effect, a thermoelectric effect and a photo-induced bolometric effect. The Fermi energy in the graphene layer determines the dominating mechanism of photocurrent generation. Assuming the Fermi energy in the range of 100 meV in our graphene flake, the effect dominating in our devices might be the photo-induced bolometric effect. As the band structure in graphene is symmetric for electrons and holes, it does not matter whether the Fermi energy is positive or negative. Most important for the detection of THz radiation is the capability to absorb photons with low energy. This is still possible via interband absorption, because at room temperature the Fermi edge is smeared out and the low frequency absorption is not blocked. More importantly intraband free-carrier absorption dominates at low energies [5]. For photocurrent generation an asymmetry of the detector is required. We note that contacting the flake with different metals for the two antenna arms did not result in an increase of the photocurrent. Apparently this provides only a negligible additional asymmetry as compared to the asymmetry provided by the flake itself.

The experimental results suggest that the detector response time is strongly affected by the substrate (cf. Fig. 2a) and 2 b)). On the low-resistive substrate the antenna forms two capacitors connected in series with SiO$_2$ as dielectric material. After charging during the FEL pulse, the capacitors are discharged via the read-out electronics and the graphene flake. The calculated capacitance of the antenna, assuming perfectly conducting plates and 300 nm of SiO$_2$ as dielectric material, is 11 pF. This results in an RC time constant of 0.6 ns, considering a load of 50 Ω. This RC time constant is in good agreement with the experimentally observed decay of the signal for the low-resistive substrate (cf. Fig. 2b)). We suggest that the same mechanism may strongly suppress the fast signal components of the detectors on the high-resistive substrate for wavelengths below 20 µm. In this case the substrate resistivity may be decreased due to thermally activated carriers caused by phonon absorption in Si. At the high average power, needed to measure the pulses with the oscilloscope, the devices strongly heat up. At a temperature of 500 K the electron density is increased to $10^{14}$ cm$^{-3}$ which corresponds to a resistivity of 40 Ω cm. This resistance may contribute to a considerable RC time constant. High-resistive substrates without absorption in the desired wavelength range, like diamond or SiC, should circumvent this problem and enable one to develop ultra-broadband detectors.

Compared to existing ultrafast detectors for the THz range working at room temperature, e.g. superlattice [14], photon drag detectors [15], or nanosize field-effect transistors [16] the detector presented in this work is easy to produce. After depositing a graphene flake to a suitable substrate only one metallization step combined with lithography is required. Although we used electron-beam lithography for our device, simpler methods like contact photolithography should also be sufficient. Furthermore the spectral range of the detector is expected to span from THz to the ultraviolet when non-absorbing substrates are used.

In conclusion we demonstrated a fast graphene-based detector working in the THz range at room temperature. The rise time of 50 ps enables the application for timing, e.g. in two-color pump-probe experiments. The presented detectors are very simple to produce, easy to handle and very robust. The saturation behavior enables one to perform autocorrelation measurements for pulse-shape analysis on a ps timescale. In addition we demonstrated the important role of the substrate and the possibility to increase the wavelength range of the detector by using different substrate materials like SiC or diamond.

We thank Peter Michel and the ELBE team for their dedicated support. Furthermore we acknowledge financial support via the Priority Programme 1459 Graphene from the German Science Foundation DFG (grant No. Wi3114/3-1 & Ga501/11-1).